%% file: main.tex
\pdfoutput=1
\documentclass[a4paper]{article}
\usepackage{INTERSPEECH2022}

\usepackage{xcolor}
\usepackage{multirow}
\usepackage{comment}
\usepackage{url}
\usepackage{cleveref}
\usepackage{multirow}

\usepackage[normalem]{ulem}

\newif\ifisdraft
\isdrafttrue 

\newcommand{\bmx}{\mathbf{X}}

\newcommand{\thetaavh}{\theta_{avh}}
\newcommand{\thetaspk}{\theta_{spk}}
\newcommand{\favh}{f_{\thetaavh}}
\newcommand{\fspk}{f_{\thetaspk}}

\definecolor{mahogany}{rgb}{0.75, 0.25, 0.0}

\ifisdraft
    \newcommand\bshi[1]{\textcolor{red}{#1 - BS}}
    \newcommand\wn[1]{\textcolor{cyan}{#1 - WN}}
    \newcommand\abdo[1]{\textcolor{olive}{#1 - ABDO}}

    \newcommand{\wnrm}[1]{\textcolor{blue}{\sout{#1}}}
    \newcommand{\bsrm}[1]{\textcolor{red}{\sout{#1}}}

\else
    \newcommand\bshi[1]{}
    \newcommand\wn[1]{}
    \newcommand\abdo[1]{}

    \newcommand{\wnrm}[1]{}
    \newcommand{\bsrm}[1]{}
\fi

\title{Learning Lip-Based Audio-Visual Speaker Embeddings with AV-HuBERT}
\name{Bowen Shi$^1$, Abdelrahman Mohamed$^2$, Wei-Ning Hsu$^2$}

\address{
  $^1$Toyota Technological Institute at Chicago \\
  $^2$Meta AI}
\email{bshi@ttic.edu, \{abdo,wnhsu\}@fb.com}

\begin{document}

\maketitle

\begin{abstract}
\input{abstract}
\end{abstract}
\noindent\textbf{Index Terms}: audio-visual, speaker verification and recognition, representation learning, self-supervised pre-training

\input{intro}
\input{method}

\input{experiment}
\input{conclusion}

\bibliographystyle{IEEEtran}
\bibliography{mybib}

\end{document}

%% file: abstract.tex
This paper investigates self-supervised pre-training for audio-visual speaker representation learning where a visual stream showing the speaker's mouth area is used alongside speech as inputs. Our study focuses on the Audio-Visual Hidden Unit BERT (AV-HuBERT) approach, a recently developed general-purpose audio-visual speech pre-training framework. We conducted extensive experiments probing the effectiveness of pre-training and visual modality. Experimental results suggest that AV-HuBERT generalizes decently to speaker related downstream tasks, improving label efficiency by roughly ten fold for both audio-only and audio-visual speaker verification. We also show that incorporating visual information, even just the lip area, greatly improves the performance and noise robustness, reducing EER by 38\% in the clean condition and 75\% in noisy conditions\footnote{Our code and models are available at \url{github.com/facebookresearch/av_hubert}.}.
%
%
%
%

%

%

%% file: intro.tex
\section{Introduction}
\label{sec:intro}

Personalizing user experiences is essential in spoken language technology systems, e.g., smart speakers and personal banking applications. Robust speaker verification (SV) and recognition models are crucial for enabling authentication and conversational experiences, as well as many other tasks like speaker diarization~\cite{Wang2018SpeakerDW}, voice conversion~\cite{Zhang2020DurIANSCDI} and source separation~\cite{Wang2019VoiceFilterTV}.

Supervised speaker representation methods made significant progress over the past decade~\cite{Snyder2018xvectors,chung2020in, desplanques2020ecapa, Lee2020two}; however, they require a non-trivial amount of human annotations of speaker identity, which might not comply with the evolving privacy-preserving standards. Furthermore, it is challenging to provide speaker labels for multi-speaker dialogues or when speakers' voices alternate between whispering and shouting~\cite{hansen2015speaker}. Self-supervised speaker representation approaches, which work around these challenges, have recently gained popularity. One family of self-supervised speaker representation methods relies on contrastive learning, which constructs positive samples by either augmenting the same speech segment or assuming a single speaker is recorded per utterance~\cite{Inoue2020SemiSupervisedCL,tao2021self,Xia2021SelfSupervisedTS}. They present solid downstream performance, where the unsupervised state-of-the-art approach~\cite{tao2021self} achieves EER of 1.66\%, close to some of the SOTA supervised systems (e.g., 0.41\% from \cite{zhao2021speakin}). However, one downside of these approaches is that they are tailored solely for speaker embedding tasks. In contrast, general self-supervised speech representation learning approaches, e.g., wav2vec 2.0~\cite{Baevski2020wav2vec2A} and HuBERT~\cite{Hsu2021HuBERT}, were found to capture enough speaker information to be competitive on SV while excelling at many other downstream tasks~\cite{Yang2021SUPERBSP}. 

The Audio-Visual Hidden Unit BERT (AV-HuBERT) was recently introduced as a general audio-visual representation learning approach. It learns joint representations over speech and lip-movement streams by alternating between clustering representations using a small codebook mimicking broad phonetic units and learning latent contextual representations through the masked prediction loss. AV-HuBERT achieves SOTA results on lip-reading and audio-visual speech recognition (AVSR) under adverse noise conditions~\cite{shi2022robust,avhubert}, thanks to the noise-immune visual modality.

This paper goes beyond single modality speaker representations to work with audio and lip-movement information to learn noise-robust speaker embeddings. We extend the representation learned by the AV-HuBERT approach to study their effectiveness for speaker-based downstream tasks in multi-modal settings.
Compared to recent specialized unsupervised speaker representation methods~\cite{Inoue2020SemiSupervisedCL,tao2021self,Xia2021SelfSupervisedTS}, one advantage of utilizing a general approach like AV-HuBERT is its ability to simultaneously serve other downstream tasks beyond speaker embedding. Prior work on audio-visual speaker representation learning focused on the consistency between the audio and the visual information, either by learning speaker embedding via audio-visual synchronization and identity matching~\cite{Nagrani2020DisentangledSE} or by multi-way matching in a joint audio-visual embedding space~\cite{Chung2020PerfectMS}. AV-HuBERT offers a more stable training procedure than methods utilizing contrastive and consistency objectives since its masked prediction loss is computed over offline learned discrete units. 

In our experiments, AV-HuBERT representations are used either in an ELMo-style feature combination protocol~\cite{sarzynska2021detecting} or through fine-tuning the whole network for the target downstream task. We report our results on speaker classification and verification tasks under four types of interfering noise and five different signal-to-noise ratios (SNR). Our audio-visual models improve label efficiency by 10 folds from supervised models, and offer 38\% and 75\% relative equal error rate (EER) reduction for SV under clean and noisy conditions compared to audio-only pre-trained models.

%
%

%
%

%
%
%
%
%

%

%

%% file: method.tex
\vspace{-0.1in}
\section{Method}
\label{sec:method}
\begin{figure}[t]
\vspace{-0.1in}
    \centering
        \caption{\label{fig:avsv-model}AV-HuBERT for learning speaker embedding. Dashed box: added during fine-tuning.}
    \includegraphics[width=\linewidth]{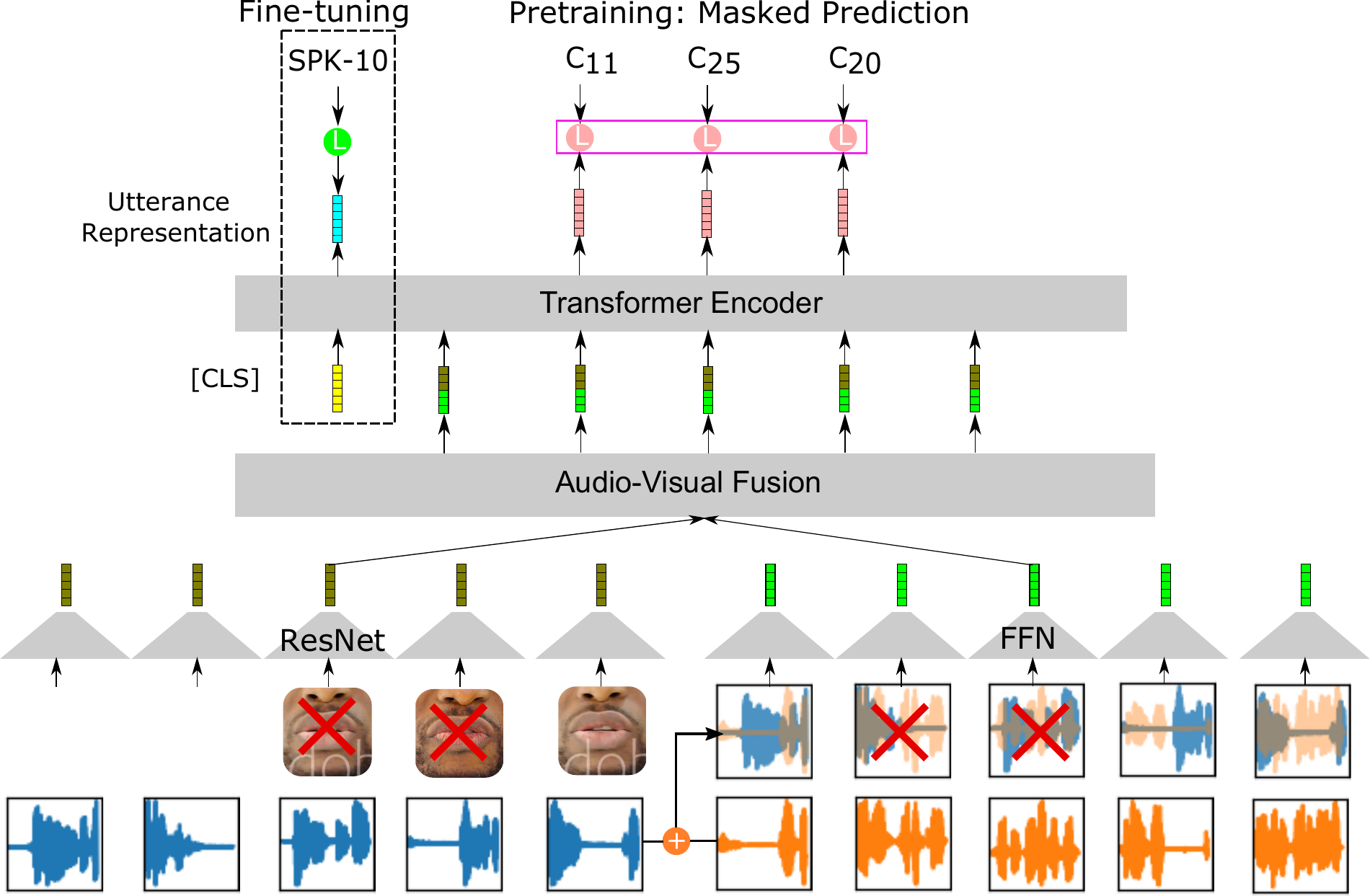}
    \vspace{-0.1in}
\end{figure}

\subsection{Overview of AV-HuBERT}
Audio-Visual Hidden Unit BERT (AV-HuBERT) is a self-supervised model that learns from unlabeled audio-visual speech data. Similar to its audio counterpart --- HuBERT~\cite{Hsu2021HuBERT}, AV-HuBERT was initially benchmarked on speech recognition tasks and achieved state-of-the-art performances on uni-modal (audio-only and video-only)~\cite{avhubert} and multimodal (audio-visual)~\cite{shi2022robust} setups. As depicted in Figure~\ref{fig:avsv-model}, AV-HuBERT comprises four modules: a feed-forward network (FFN) audio feature extractor, a modified ResNet~\cite{Stafylakis2017CombiningRN,Martnez2020LipreadingUT} video feature extractor, a fusion module, and a Transformer~\cite{Vaswani2017attention} backend. The two feature extractors generate frame-level representation for the corresponding stream, which are frame-wise concatenated by the fusion module to form initial audio-visual features. The Transformer backend takes these features and produces the contextualized frame-level audio-visual representations. The entire model is optimized to perform masked prediction, where random segments are masked for each stream independently (denoted by the red crosses in Figure~\ref{fig:avsv-model}), and the model learns to predicts the cluster assignment of the masked frames (the middle three frames in the example). The cluster assignment are iteratively refined: it is produced by clustering MFCC features in the first iteration, and by clustering previous iteration's AV-HuBERT representations in the subsequent iterations. 

While the self-supervised learning objective of HuBERT and AV-HuBERT (masked prediction of cluster assignments) resembles the task of automatic speech recognition (ASR)~\cite{avhubert}, several analyses observed that such models still learn rich speaker information especially in earlier layers~\cite{chang2021distilhubert, Yang2021SUPERBSP, Chen2021WavLM}.
These observations can be linked to many studies in the ASR and SV literature. For example, researchers have found that speaker information can improve ASR performance
by encoding the effects to which an ASR system should be invariant~\cite{senior2014improving}.
Hence, AV-HuBERT may learn to extract speaker information at early layers to better infer phonetic information in the subsequent layers. In addition, statistics of the acoustic features extracted from ASR models have been widely used for SV~\cite{lei2014novel}, which can also explain why HuBERT features benefits SV. Motivated by the strong results of adapting HuBERT for SV, we explore learning audio-visual speaker embeddings from a pre-trained AV-HuBERT model with emphasis on the noise robustness aspect strengthened by the addition of visual stream.

\begin{figure}[t]
\vspace{-0.1in}
    \centering
        \caption{Example lip images, from LRS3 dataset~\cite{afouras2018lrs3} \label{fig:mouth-grid}. Images in the same row are from the same speaker. }
        \vspace{-0.1in}
    \includegraphics[width=0.8\linewidth]{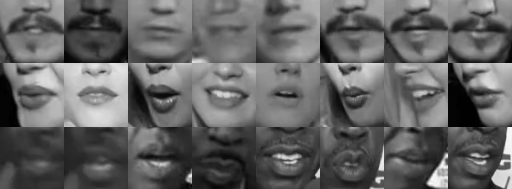} %
    \vspace{-0.2in}
\end{figure}

\subsection{Learning multimodal speaker embedding}

We describe in this section how to learn speaker embeddings with pairs of audio-visual speech $\bmx$ and speaker label $y$.
Given a pre-trained AV-HuBERT model $\favh$ parameterized by $\thetaavh$, the goal is to leverage it to learn a speaker embedder $\fspk$ such that $\fspk(\bmx_i)$ are similar to $\fspk(\bmx_j)$ if $y_i = y_j$, and are dissimilar otherwise.

Two learning protocols are considered, which differs in whether $\favh$ is frozen or not. 
The former is used to evaluate representation quality by treating $\favh$ as a fixed feature extractor. 
Following~\cite{Yang2021SUPERBSP}, we consider ELMo~\cite{Peters2018DeepCW} style fine-tuning, where frame-level AV-HuBERT features at each Transformer layer $\mathbf{C}^{(l)} = \favh^{(l)}(\bmx) \in \mathbb{R}^{D_{avh} \times T}$ are weighted summed using learnable non-negative weights $w^{(l)} \in \mathbb{R}_{\ge 0}$ that sums to one: $\sum_{l=1}^L w^{(l)} = 1$. This representation is then passed to the downstream model $f_{\theta_{d}}$ to produce a fixed-sized speaker embedding: $\bm{s} = f_{\theta_{d}} (\sum_{l=1}^L w^{(l)} \mathbf{C}^{(l)}) \in \mathbb{R}^{D_{spk}}$. Only $w^{(l)}$ and $\theta_{d}$ are updated during fine-tuning. In this protocol, we adopt the same downstream models in~\cite{Yang2021SUPERBSP}.

The other protocol evaluates pre-training~\cite{bert, Baevski2020wav2vec2A, Hsu2021HuBERT, baevski2022data2vec}, where $\theta_{avh}$ is updated along with new parameters during fine-tuning. We prepend a learnable \texttt{[cls]} embedding $\mathbf{e}_{cls} \in \mathbb{R}^{D_{avh}}$ to the input sequence of the Transformer encoder $\mathbf{Z} \in \mathbb{R}^{D_{avh} \times T}$. The augmented input $\mathbf{\tilde{Z}} \in \mathbb{R}^{D_{avh} \times (T+1)}$ is then passed to the Transformer encoder to produce $\mathbf{\tilde{C}}^{(L)} \in \mathbb{R}^{D_{avh} \times (T+1)}$ at the final layer $L$. The contextualized embedding corresponding to the \texttt{[CLS]} position, $\mathbf{\tilde{C}}^{(L)}_{cls} \in \mathbb{R}^{D_{avh}}$ serves as the speaker embedding $\bm{s}$.

For both protocols, we append a softmax layer that takes a speaker embedding as input and outputs the posterior $softmax(\mathbf{A} \bm{s}) \in [0, 1]^{C}$ over $C$ training set speakers. Learnable parameters along with the softmax layer are fine-tuned to minimize a cross entropy loss with respect to speaker labels.
%
\vspace{-0.1in}

\subsection{Trade-off between lip and face-based visual input}

Visual input (lip or face video) can improve speaker embeddings on top of audio input from two aspects: first, it provides appearance features that is complementary to voice features, resulting in gain in clean and noisy conditions; second, it helps anchor the target speech from a mixture of sounds, making embeddings more resilient to noise. Most prior work use whole-face videos as visual input~\cite{shon2019noise, Sari2021AMA}. While it can in principle lead to better performance, such a system also harvests more potentially sensitive data. 
In this work, we present a lip-based system (see Figure~\ref{fig:mouth-grid}) that enjoys superior noise robustness (Tabel~\ref{tab:noise-aug}) while demanding less information. The accuracy trade-off is studied in Section 3.4.

%% file: experiment.tex
\section{Experiments}
\label{sec:exp}

\subsection{Setup}\label{sec:setup}

\noindent 
\textbf{Pre-training}\hspace{.5em} We pre-training a 12-layer BASE and 24-layer LARGE AV-HuBERT following~\cite{avhubert} and \cite{shi2022robust} with only one change. The dimension of features from BASE and LARGE model are 768 and 1,024 respectively. In~\cite{avhubert}, models are pre-trained with 433 hours of LRS3~\cite{afouras2018lrs3} and the English portion of VoxCeleb2 (VC2)~\cite{voxceleb2} from both the dev and test split. In this paper, since we evaluate the speaker embeddings on the VC2 test split and it contains both English and non-English speakers, we combine LRS3 with all VC2 data, English or non-English, but exclude the VC2 test split for pre-training, which sums up to roughly 2,800 hours. Both LRS3 and VC2 are sampled at a frame rate of 25Hz. The AV-HuBERT model also produces representations at this frame rate. To improve noise robustness, noise randomly sampled from MUSAN~\cite{Snyder2015MUSANAM} is added to the audio stream following~\cite{shi2022robust}. The audio and video preprocessing steps remain the same as~\cite{avhubert}.

\vspace{2pt}
\noindent 
\textbf{Fine-tuning}\hspace{.5em} 
For the frozen protocol,
to compare with results reported in~\cite{Yang2021SUPERBSP}, we adopt the same prediction heads as~\cite{Yang2021SUPERBSP}, which is an average pooling layer for SC, and an x-vector model~\cite{snyder2018x} for SV. 
Two widely used audio-visual speech recognition datasets, VoxCeleb1 (VC1, 352 hours/1,251 speakers)~\cite{nagrani2017voxceleb}\footnote{
The video files from~\cite{Nagrani18seeing} was down-sampled by a factor of six ($\sim$4.17Hz). To tackle this, we let the ResNet video feature extractor to process the downsampled video as is and upsample its output by a factor of six before passing it to the Transformer. Empirically, this leads to similar performance as upsampling the video at input to the ResNet while reducing the memory and the compute. 
} and VC2 (2,442 hours/6,112 speakers), are adopted for supervised fine-tuning. Noise-augmented fine-tuning with MUSAN following~\cite{shi2022robust} is explored in \S\ref{sec:noise_aug}.
All models are optimized with Adam~\cite{kingma2014adam} with learning rate warmed up to 0.001 for one third of training steps and then linearly decayed. The pre-trained parameters are frozen for certain steps before being updated. For \{5h, 50h, 500h, VC2\} setups in Table~\ref{tab:lab-eff}, we train the model for \{20, 30, 90, 75\} K-steps with a batch size of \{100, 100, 100, 400\} and \{60, 60, 120, 240\} for audio-only and audio-visual setting.
Whenever the model is used in audio-only setting, the visual feature (ResNet output) is replaced by an all-zero vector.

\vspace{2pt}
\noindent
\textbf{Evaluation}\hspace{.5em}
Two tasks from \cite{Yang2021SUPERBSP}, speaker verification (SV) and speaker classification (SC), are used. 
SC evaluates prediction accuracy of a closed set speaker seen during training. We report speaker accuracy (Acc) on VC1 using the official speaker identification split. 
SV considers an speaker-independent setup, where a set of testing trials is provided, each of which contains two utterances and a label indicating if the two are from the same speaker. 
We report EER on VC1 using its official test trials (37,611 trials) and on VC2 by sampling one positive trial and one negative trial for each test set utterance (72,474 trials). In terms of trial scoring, we adopt the protocol used in~\cite{chung2020in}: for any pair of audios $\bm{X}_i$ and $\bm{X}_j$, ten 4-second segments ($\bm{X}_{i,n}$ and $\bm{X}_{j,m}$ for $n, m \in 1, \cdots, 10$) are evenly sampled from each audio and we use average cosine similarity of the segment speaker embeddings ($\bm{s}_{i, n}$ and $\bm{s}_{j, m}$) between all combinations, $avg(\sum_{n=1}^{10} \sum_{m=1}^{10} cos(\bm{s}_{i, n}, \bm{s}_{j, m}))$, as their matching score.

%

%


%
To probe noise robustness, we follow \cite{shi2022robust} to create 20 noisy test sets for VC1 and VC2, where each clean set is mixed with \{Babble, Speech, Music, Other\} noise at a SNR in \{-10, -5, 0, 5, 10\} dB.
By default, we fine-tune all models on VC2 without noise augmentation and with the protocol where AV-HuBERT parameters are updated, and report EER on the clean set. We use AV-HuBERT Base for analysis purpose (sec~\ref{sec:eff_pt}-\ref{sec:face}).

\vspace{-0.1in}
\subsection{Effectiveness of AV-HuBERT pre-training}\label{sec:eff_pt}

\input{tables/lab-eff}

We first study label efficiency of AV-HuBERT pre-training for audio-only and audio-visual speaker verification, and evaluate their performance on both the clean and noisy test trials (Table~\ref{tab:lab-eff}). 
In the ``noisy'' columns, we report the average EER over the 20 testing configurations.
Three sizes of subsets are considered: 20\%/2\%/0.2\% of the labeled data. For each size, we generate subsets in two ways: one is by sampling x\% of the utterances, and the other is by sampling x\% of the speakers and select all their utterances. For a given size, the former would contain more speakers but less utterances per speaker. 

For each labeled subset, AV-HuBERT outperforms models trained from scratch (PT=None) when the same modalities are used as input. Comparing across subsets, we find AV-HuBERT often matches or outperforms a randomly initialized model that use 10 times more data
(e.g., AV-HuBERT is 16.7\% on VC2 clean set with VC2-5h data and AV input, while the baseline is 21.5\% with VC2-50h data and AV input).

Next, we observe that with pre-training, audio-visual models always outperform audio-only ones. The consistent gain on the clean set shows that lip videos bring complementary information for recognizing speakers, which is different from speech recognition where visual information helps very little in clean conditions~\cite{shi2022robust,ma2021conformer}. On the other hand, audio-visual based models bring substantial gains in noisy conditions compared to audio-based models (e.g., reducing EER from 20.0\% to 2.5\% on noisy VC1 test set when pre-trained and fine-tuned on the entire VC2). This demonstrates the enhancement in noise robustness by incorporating lip information.

Comparing the different subset sampling strategies for 0.2\% VC2 (VC2-5h versus VC2-15spk), it shows that pre-trained models benefit from having more speakers (w/ PT, 16.7\% for VC2-5h vs. 22.6\% for VC2-15spk on VC2 clean with AV input),\footnote{
The model achieves similar performance on noisy test sets for both subset sampling strategies, which is caused by the mismatch in training/testing audio conditions (clean vs. noisy). With noise-augmented fine-tuning, the same trend emerges again (w/ PT, 5h vs. 15spk: 26.1\%/13.8\% vs. 30.7\%/21.6\% with A/AV input on VC1 noisy).
} while the supervised audio-visual baselines struggle with too few examples per speaker (w/o PT 32.8\% vs. 29.8\% on VC2 clean with AV input). 
This implies that a visual encoder trained from scratch hardly benefits from increasing number of speakers given only few utterances per speaker.
In the 5h-setting (few utterances per speaker), the extra visual modality is only effective in case of pre-training whereas an audio-visual model trained from scratch is consistently worse than its audio-only counterpart regardless of auditory conditions (PT=None, A vs. AV). To conclude, we discover pre-training can lead to better generalization with few-shot learning.


%

%

%
%

%

%

%

%
%

%

%

\subsection{Noise-augmented fine-tuning}\label{sec:noise_aug}
Observing the gap between audio-based and audio-visual models on SV in the previous section, we study if applying noise augmentation during fine-tuning can bridge the gap. We fine-tune AV-HuBERT on VC2-500h audio and audio-visual data with and without noise augmentation described in \S\ref{sec:setup}

The performance of the four models (A or AV, with or without noise augmentation) on the 20 noisy test sets are presented in Table~\ref{tab:noise-aug}. Unsurprisingly, all models perform worse on lower SNR conditions. Nevertheless, we observe much bigger degradation for audio-based model, especially when corrupted with speech (S) or babble (B) noise, because audio models can not determine who the target speaker is from a mixture of speech.
In contrast, audio-visual based models suffer very minor performance degradation in noisier conditions, because lip videos can help identify the target to infer speaker embedding for.

We also see noise augmentation reduces the average EER at -10dB from 43.7\% to 30.8\% for audio-only models, and from 6.3\% to 3.3\% for audio-visual models. The results suggest that while noise-augmentation is beneficial, adopting it alone can not bridge the gap between audio and audio-visual models pre-trained and fine-tuned on the same amount of data.

\input{tables/noise-aug}

\subsection{Choice of visual input}\label{sec:face}
Using lip instead of face videos is a key feature of the proposed model as it better addresses data input concerns while exhibits excellent noise robustness. We quantify how much our model degrades by comparing it with a variant of AV-HuBERT pre-trained and fine-tuned on face videos. As a baseline, we also evaluate a widely used face recognition model RetinaFace~\cite{deng2020retina} taking a still image as input, which only achieves an EER of 14.6\% on VC2. Table~\ref{tab:face-vs-mouth} shows that the face-based AV-HuBERT model indeed performs slightly better (0.9\% absolute EER reduction) , which is a trade-off to be considered.

\input{tables/face-lip}

\input{tables/overall-comparison}

\subsection{Comparison with prior work}\label{sec:prior}
We first compare AV-HuBERT with other self-supervised models using the SUPERB~\cite{Yang2021SUPERBSP} protocol that evaluates the quality of frozen representations (upper half of Table~\ref{tab:cmp_all}), where models are fine-tuned and evaluated on VC1 for SC and SV. 
With audio-only input, AV-HuBERT-L outperforms wav2vec2-L and HuBERT-L and is inferior to WavLM-L. 
Outperforming all audio-only pre-trained models, AV-HuBERT achieves significantly better results on the SC and SV tasks with audio-visual input even with an order of magnitude less pre-training data. 
%

The lower half of Table~\ref{tab:cmp_all} compares AV-HuBERT with other supervised and self-supervised models reporting results on VC1 or VC2 without following the SUPERB protocol.
Specifically, we fine-tune our pre-trained model using the whole VC2 labeled data. As is expected, fine-tuning with a large amount of labeled data improves performance. In audio-visual setting, our best model (0.84\%) outperforms~\cite{Sari2021AMA} (1.8\% and 1.4\%) with a single model and slightly falls behind its ensembled model (0.7\%). Note in contrast to the prior works~\cite{Nagrani2020DisentangledSE,Sari2021AMA,shon2019noise} which uses the whole face, our model only relies on the lip area of the speaker as visual input and achieves a better trade-off between required information and performance. In addition, we acknowledge the gap between our best model and the current SOTA on VC1 (\cite{Chen2021WavLM}: 0.38\%). Such gap is attributed to the relative smaller amount of pre-training data we use as well as the simplicity of our training and evaluation pipeline (e.g., \cite{Chen2021WavLM} uses Inter-TopK penalty~\cite{zhao2021speakin}, ECAPA-TDNN~\cite{desplanques2020ecapa} as an upstream network, second stage large-margin fine-tuning, adaptive s-norm~\cite{Karam2011TowardsRF,Cumani2011ComparisonOS} and calibration~\cite{Thienpondt2021theidlab} in evaluation, etc). As the goal of this paper is to study the speaker embedding learned by AV-HuBERT, we stick to a simple and standard pipeline for speaker verification. 
How to incorporate extra techniques in AV-HuBERT framework is left as our future work. 
%

%
%

%
%

%
%
%
%
%
%

%% file: tables/lab-eff.tex
\begin{table}[htp]
    \centering
    \caption{\label{tab:lab-eff}
    SV performance on clean and noisy test sets when fine-tuned with various VC2 subsets. The EER averaged over 20 setups (5 SNRs $\times$ 4 types) is reported for the noisy test sets.
    }
    \vspace{-0.1in}
    \resizebox{\linewidth}{!}{
    
    \begin{tabular}{ccc|cccc}
        \multirow{2}{*}{PT} & \multirow{2}{*}{FT} & \multirow{2}{*}{Mod.} 
        & \multicolumn{2}{c}{VC2 EER (\%)} & \multicolumn{2}{c}{VC1 EER (\%)} \\
        & & & clean & noisy & clean & noisy \\
        \midrule
        \midrule

        None        & \multirow{4}{*}{\shortstack{VC2-15spk\\({5h})}} & A  & 26.8 & 39.2 & 25.1 & 39.2 \\
        None        &  & AV & 29.8 & 35.9 & 24.6 & 28.7 \\
        VC2+LRS3    &  & A  & 23.3 & 33.9 & 20.0 & 33.0 \\
        VC2+LRS3    &  & AV & 22.6 & 28.0 & 19.4 & 21.9 \\
        \midrule
        None        & \multirow{4}{*}{\shortstack{VC2-156spk\\({50h})}} & A  & 18.5 & 34.5 & 16.1 & 34.6 \\
        None        &  & AV & 16.4 & 24.7 & 13.1 & 17.7 \\
        VC2+LRS3    &  & A  & 11.8 & 28.9 & 9.4 & 29.1 \\
        VC2+LRS3    &  & AV & 9.3 & 18.8 &  7.8 & 12.5 \\

        \midrule
        None        & \multirow{4}{*}{\shortstack{VC2-1200spk\\({485h})}} & A  & 11.1 &  31.6 & 8.6 &  30.5 \\
        None        &  & AV &  9.3 &  17.6 & 7.0 &  9.9 \\
        VC2+LRS3    &  & A  &  7.2 &  26.1 & 4.9 &  25.2 \\
        VC2+LRS3    &  & AV &  5.7 &  12.6 & 3.8 & 6.1 \\
        \midrule
        \midrule
        None        & \multirow{4}{*}{\shortstack{VC2-5h\\({1740spk})}} & A  & 24.4 &  39.4 & 21.7 &  39.5 \\
        None        &  & AV & 32.8 &  41.0 & 30.2 & 40.3 \\
        VC2+LRS3    &  & A  & 20.1 &  34.7 & 17.7 &  34.5 \\
        VC2+LRS3    &  & AV & 16.7 &  28.6 & 13.9 & 23.0 \\
        \midrule
        None        & \multirow{4}{*}{\shortstack{VC2-50h\\({5113spk})}} & A  & 20.2 &  35.5 & 16.1 &  34.7 \\
        None        &   & AV & 21.5 &  26.3 & 15.7 &  16.4 \\
        VC2+LRS3    &  & A  & 10.7 & 29.7 &  8.0 &  28.7\\
        VC2+LRS3    &  & AV &  7.4 &  19.8 &  4.8 &  11.4\\
        \midrule
        None        & \multirow{4}{*}{\shortstack{VC2-500h\\({5992spk})}} & A  & 10.6 & 33.1 & 8.0 & 31.4 \\
        None        &  & AV &  6.5 & 14.5 & 5.3 &  7.8 \\
        VC2+LRS3    &  & A  &  4.9 & 23.7 & 3.0 & 22.8 \\
        VC2+LRS3    &  & AV &  3.7 &  9.2 & 1.7 &  3.9 \\
        \midrule
        \midrule
        None        & \multirow{4}{*}{\shortstack{VC2\\({5994spk})}}      & A  &  7.3 & 29.2 & 5.1 & 27.8 \\
        None        &       & AV &  5.1 & 11.3 & 2.9 &  4.7 \\
        VC2+LRS3    &      & A  &  3.4 & 20.9 & 1.9 & 20.0 \\
        VC2+LRS3    &      & AV & 2.4  &  7.8 & 1.0 &  2.5 \\
    \end{tabular}
    }
    \vspace{-0.2in}
\end{table}

%% file: tables/noise-aug.tex
\vspace{-0.1in}

\begin{table}[h]
    \centering
    \caption{AV-HuBERT fine-tuned on VC2-500h with audio (A) or audio-visual (AV) input, with or without noise augmentation. The abbreviations used are B: Babble, S: Speech, M: Music, and O: Other.\label{tab:noise-aug}}
    \vspace{-0.1in}
    \resizebox{\linewidth}{!}{
    \begin{tabular}{cc|ccccc|ccccc}
    Noise & Noise & \multicolumn{5}{c|}{A, VC1 EER (\%), SNR (dB)=} & \multicolumn{5}{c}{AV, VC1 EER (\%), SNR (dB)=} \\
    Aug?  & Type  &  -10 & -5 & 0 & 5 & 10 & -10 & -5 & 0 & 5 & 10 \\
    \midrule
    \multirow{4}{*}{N}
    & B & 48.2 & 36.4 & 18.5 & 9.6  & 6.0 &  4.4 & 3.9 & 3.4 & 2.6 & 2.2 \\
    & S & 48.8 & 46.5 & 36.5 & 18.3 & 8.5 &  8.6 & 6.8 & 4.8 & 3.4 & 2.5 \\
    & M & 39.3 & 26.9 & 14.5 & 8.3  & 5.5 &  6.2 & 4.3 & 3.1 & 2.4 & 2.0 \\ 
    & O & 34.0 & 23.3 & 13.7 & 8.8  & 5.9 &  6.0 & 4.3 & 3.2 & 2.6 & 2.3 \\
    \midrule
    \multirow{4}{*}{Y}
    & B & 48.1 & 27.2 & 12.7 & 7.3  & 5.2 &  3.4 & 3.2 & 2.5 & 2.2 & 2.0 \\
    & S & 24.4 & 14.9 & 11.8 & 12.3 & 9.6 &  3.2 & 2.8 & 2.6 & 2.3 & 2.0 \\
    & M & 27.3 & 14.3 & 8.2  & 5.6  & 4.4 &  3.5 & 2.8 & 2.4 & 2.0 & 1.8 \\
    & O & 23.6 & 13.0 & 8.0  & 5.8  & 4.7 &  3.1 & 2.6 & 2.3 & 2.1 & 2.0 \\

    \end{tabular}
        }
\end{table}
\vspace{-0.2in}

%% file: tables/face-lip.tex
\vspace{-0.1in}
\begin{table}[htp]
    \centering
    \setlength{\tabcolsep}{3pt}
    \caption{\label{tab:face-vs-mouth}
    Comparing different input. AV-HuBERT is fine-tuned on VC2-500h.}
    \vspace{-0.1in}
    \begin{tabular}{cc|c}
        Model & Input & VC2 EER (\%) \\
        \midrule
        AV-HuBERT & audio + face video & 2.8 \\
        AV-HuBERT & audio + lip video & 3.7 \\
    \end{tabular}
    \vspace{-0.2in}
\end{table}

%% file: tables/overall-comparison.tex
\begin{table}[ht]
    \centering
    \caption{(Top) Comparison with the prior work following the SUPERB fine-tuning protocol. Models are fine-tuned on VC1. (Bottom) Comparison with prior work that does not follow the SUPERB evaluation protocol.}\label{tab:cmp_all}
    \vspace{-0.1in}
    \resizebox{\linewidth}{!}{
    \begin{tabular}{ccc|cc}
        \multirow{2}{*}{Method} & PT & \multirow{2}{*}{Mod.} & \multicolumn{2}{c}{VC1} \\
        & Data & & SC-Acc & SV-EER  \\ 
        
        \midrule
        FBANK~\cite{Yang2021SUPERBSP} & - & A & 8.5E-4 & 9.56 \\
        wav2vec2-B~\cite{Yang2021SUPERBSP} & LS (960 hr) & A & 75.18 & 6.02 \\
        HuBERT-B~\cite{Yang2021SUPERBSP} & LS (960 hr) & A & 81.42 & 5.11 \\
        WavLM-B~\cite{Chen2021WavLM}  & Mix (94k hr) & A & 89.42 & 4.07 \\
        wav2vec2-L~\cite{Yang2021SUPERBSP} & LL (60k hr) & A & 86.14 & 5.65 \\
        HuBERT-L~\cite{Yang2021SUPERBSP} & LL (60k hr) & A & 90.33 & 5.98 \\
        WavLM-L~\cite{Chen2021WavLM}  & Mix (94k hr) & A & 95.49 & 3.77 \\
        \midrule
        AV-HuBERT-B & \multirow{4}{*}{\shortstack{VC2+LRS3 \\(2.8k hr)}} & A  & 80.99 & 5.85 \\
        AV-HuBERT-B &  & AV & 93.90 & 4.85 \\
        AV-HuBERT-L &  & A  & 91.56 & 4.42 \\
        AV-HuBERT-L &  & AV & 98.06 & 2.95 \\
    \end{tabular}
    }
    
    \vspace{1em}
    
    \resizebox{\linewidth}{!}{
    \begin{tabular}{cccc|ccc}
        \multirow{2}{*}{Method} & PT & FT & \multirow{2}{*}{Mod.} & VC1 & VC1  & VC2 \\
        & Data & Data & & SC-Acc & SV-EER & SV-EER \\
        \midrule
        Nagrani et al.~\cite{Nagrani2020DisentangledSE} & 20\% VC2 & VC1 & AV & - & 9.43 & - \\
        WavLM-L~\cite{Chen2021WavLM} & Mix (94k hr) & VC2 & A & - & 0.38 & - \\
        Shon et al.~\cite{shon2019noise} & - & VC2 & AV & - & - & 5.29 \\
        Unimodal~\cite{Sari2021AMA} & - & VC2 & A  & - & 2.2 & 3.5 \\
        Multi-view~\cite{Sari2021AMA} & - & VC2 & AV & - & 1.8 & 2.4 \\
        Feature Fusion~\cite{Sari2021AMA} & - & VC2 & AV  & - & 1.4 & 2.0 \\
        Ensemble~\cite{Sari2021AMA} & - & VC2 & AV  & - & 0.7 & 1.6 \\
        \midrule
        AV-HuBERT-B & \multirow{4}{*}{\shortstack{VC2+LRS3 \\(2.8k hr)}} & VC2 & A  & -     & 1.92  & 3.43  \\
        AV-HuBERT-B &  & VC2 & AV & -     & 1.00  & 2.41  \\
        AV-HuBERT-L &  & VC2 & A  & -     & 1.71  & 3.11 \\
        AV-HuBERT-L &  & VC2 & AV & -     & 0.84  & 2.29 \\

    \end{tabular}
     }
     \vspace{-0.2in}
\end{table}

%% file: conclusion.tex
\section{Conclusion}
\label{sec:conclusion}

In this paper, we study learning lip-based audio-visual speaker embeddings from AV-HuBERT. We show that the general-purpose audio-visual speech representation learning framework, AV-HuBERT, is able to learn high-quality speaker embeddings that improves the performance on speaker related tasks including classification and verification using less labeled data. The proposed model also greatly improves robustness to a variety of noise while requiring only lip videos instead of whole face videos.